\documentclass{article}
\newcommand{\bfr}{\begin{flushright}}
\newcommand{\efr}{\end{flushright}}
 
\begin{document}
\title{Wilson Loops in Open String Theory
}
\author{Kiyoshi Shiraishi\\
Department of Physics, Tokyo Metropolitan University,\\ Setagaya, Tokyo,
158, Japan
}
\date{Modern Physics Letters {\bf A3} (1988) pp. 283--287
}
\maketitle
\begin{abstract}
Wilson loop elements on torus are introduced into the partition
function of open strings as Polyakov's path integral at one-loop
level. Mass spectra from compactification and expected symmetry
breaking are illustrated by choosing the correct weight for the
contributions from annulus and M\"obius strip. We show that Jacobi's
imaginary transformation connects the mass spectra with the Wilson
loops. The application to thermopartition function and cosmological
implications are briefly discussed.
\end{abstract}

The string theory \cite{1} as a candidate for a unified model of
interactions contains gauge fields with a large symmetry.  The gauge
symmetry must be broken down to more or less ``realistic'' symmetry,
such as $SU(3)\times SU(2)\times U(1)\times\cdots$, in an early stage
of the universe. A symmetry breaking mechanism has been proposed, which
is called the Wilson loop mechanism.\cite{2} Roughly speaking, it can
be said that the gauge fields with non-zero vacuum expectation values
(up to gauge transformations) on a non-simply connected space play the
role of Higgs fields in a field theoretical perspectlve.

Some simple models were investigated in the framework of Kaluza-Klein
theory.\cite{3}  An important point is that the energies of vacua
corresponding to different symmetries are degenerate at classical
level.\cite{3,4} The vacuum energy from one-loop quantum effect must be
taken into consideration in order to determine the ``true'' vacuum. The
calculation was made at zero temperature as well as finite
temperature.\cite{5,6} In order to evaluate the free energies in
Kaluza-Klein theory, we need to know the excitation modes which arise
from the compactification of the extra space. In the presence of the
vacuum gauge fields, such mass spectra might shift up or down according
to the magnitude of the vacuum gauge fields and the coupling to the
field. The vacuum and free energies are obtained by using the mass
specra and they are expressed as a function of the vacuum gauge fields.

On the other hand, in string theories, the vacuum energy or
cosmological constant is computed in Polyakov's path integral method
most clearly.\cite{7} The free energy at finite temperature can also be
obtained in path integral form, as is shown by Polchinski in the case
of bosonic strings.\cite{7} The path integral approach is also useful
for investigation of torus compactifications with background
fields.\cite{8} 

In this paper, we will show how the vacuum gauge fields on torus
modify the mass spectra in open string theory in terms of the
path integral method. Polyakov's path integral formalism for the
open string is developed by several authors.\cite{9} For our
purpose, we have only to consider the zero-mode pieces of bosonic
string coordinates and the background gauge fields. In other
words, the ``Kaluza-Klein'' excitation is added to each
``stringy'' mass level. Since the generalization to the
superstring is straightforward, we start with the following world-sheet
action for the bosonic strings, which was given by Callan et
al.\cite{10} as
\begin{equation}
S=\frac{T}{2}\int
d^2\sigma\sqrt{g}g^{ab}\partial_aX^M\partial_bX_M+iT\oint ds\,
A_M\frac{\partial}{\partial s}X^M\,,
\label{1}
\end{equation}
where $X^M$ are bosonic fields and $T$ is the string tension. The line
integral is assumed to be path-ordered on the boundary of the world
sheet. We can take the world-sheet metric $g_{ab}$ for annulus, such
that
\begin{equation}
d^2\sigma=g_{ab}d\sigma^a
d\sigma^b=d\sigma_1^2+t^2 d\sigma_2^2\qquad (0\le\sigma_1\le 1,~
0\le\sigma_2\le 1)\,, 
\end{equation}
where $t$ is the moduli parameter. (For M\"obius strip and other
configurations, parametrization of $g_{ab}$ is given in Ref. \cite{9}.)

From now on, we only consider the world sheet configurations which have
boundaries, since the gauge field is attached to boundaries. The traces
of two ends of an open string correspond to boundaries of the world
sheet at $\sigma_1=0$ and $\sigma_1=1$.

For simplicity, we consider $SO(N)$ as the gauge group and a torus in
one space dimension. Then $A_M$ is $N\times N$ matrix valued.

In this compact dimension, denoted by $I$-th dimension, string
coordinate field can be written as
\begin{equation}
X^I=x^I+2\pi r\ell\sigma_2+({\rm oscillators})\,,
\label{3}
\end{equation}
where $\ell$ is an integer. The oscillator part is expressed in a
function periodic in $\sigma_2$.

We set the radius of the torus to $r$ in Eq.~(\ref{3}). This means that
the points separated by $2\pi r$ in the $I$-th direction are identified
to a point on the torus.

We find that substitution of the zero-mode piece in Eq.~(\ref{3}) into
the action (\ref{1}) yields the Wilson loop element; the integer $\ell$
indicates the winding number of the Wilson line around the torus.

The partition function in the path integral form is proportional to the
factor
\begin{equation}
f=\sum_{\ell=-\infty}^\infty \exp\left[-\frac{(2\pi
r)^2T\ell^2}{2t}\right]\prod_{boundaries}{\rm Tr~}\exp(i2\pi
r\ell A_IL_B)\,,
\end{equation}
where $L_B$ is the length of the boundary. An annulus has two boundaries
with $L_B=1$ while a M\"obius strip has one boundary with $L_B=2$.

Now we study the mass spectra which come from the compactification and
the nature of the symmetry breaking induced by $A_I$. For concreteness,
we take $A_I$ as the following $N\times N$ matrix
\begin{equation}
A_I=A\left[\begin{array}{cccc}
0 & -i & 0 &\cdots\\
i & 0 & 0 &\cdots\\
0 & 0 & 0 & \\
\vdots & \vdots & &\ddots
\end{array}\right]\,.
\end{equation}
Then we get the factors for an annulus and a M\"obius strip respectively
as:
\begin{eqnarray}
& &\mbox{For an annulus,}\qquad\nonumber \\
& &f_A=\sum_{\ell=-\infty}^\infty \exp\left[-\frac{(2\pi
r)^2T\ell^2}{2t}\right]\cdot [N+2\{\cos(2\pi\phi\ell)-1\}]^2\,;\\
& &\mbox{For a M\"obius strip,}\qquad\nonumber\\
& &f_M=\sum_{\ell=-\infty}^\infty \exp\left[-\frac{(2\pi
r)^2T\ell^2}{2t}\right]\cdot [N+2\{\cos(4\pi\phi\ell)-1\}]\,,
\end{eqnarray}
where $\phi\equiv rTA$.

The weight for each contribution must be specified when we sum up these
two. We can determine it by comparing the trivial case, $\phi=0$, with
the known result.\cite{9} For the open bosonic and superstring,
massless modes of ``stringy'' excitation must contribute to the vacuum
amplitude with degeneracy $N(N-1)/2$, the degree of freedom in the
adjoint representation. The Kaluza-Klein mode is added to each stringy
excitation, which is labelled by an occupation number of oscillators.

Let us consider the excitation for each stringy mode. It is well known
that the relative sign of contributions from annulus and M\"obius strip
is alternating according to even and odd mode.

To get the Kaluza-Klein mass spectra, we use Jacobi's imaginary
transformation such as, in our case,
\begin{equation}
\sum_{\ell=-\infty}^\infty \exp\left[-\frac{(2\pi
r)^2T\ell^2}{2t}\right]\cdot \cos(2\pi\phi\ell)=
\sqrt{\frac{t}{2\pi Tr^2}}\sum_{\ell=-\infty}^\infty
\exp\left[-\frac{\pi t(\ell-\phi)^2}{2\pi Tr^2}\right]\,.
\end{equation}

Then we find the following results. For even modes, which include the
massless mode, the integrand contains the factor
\begin{eqnarray}
\frac{f_A-f_M}{2}&=&\sqrt{\frac{t}{2\pi Tr^2}}
\cdot\left[[(N-2)(N-3)/2+1]\cdot\sum_\ell\exp\left[-\frac{\pi
t \ell^2}{2\pi Tr^2}\right]\right.\nonumber \\
& &\left.+2(N-2)\cdot\sum_\ell\exp\left[-\frac{\pi
t (\ell-\phi)^2}{2\pi Tr^2}\right]\right]\,,
\end{eqnarray}
while for odd modes, it contains the factor
\begin{eqnarray}
\frac{f_A+f_M}{2}&=&\sqrt{\frac{t}{2\pi Tr^2}}
\cdot\left[[(N-2)(N-1)/2+1]\cdot\sum_\ell\exp\left[-\frac{\pi
t \ell^2}{2\pi Tr^2}\right]\right.\nonumber \\
& &\left.+2(N-2)\cdot\sum_\ell\exp\left[-\frac{\pi
t (\ell-\phi)^2}{2\pi Tr^2}\right]\right.\nonumber \\
& &\left.+2\cdot\sum_\ell\exp\left[-\frac{\pi
t (\ell-2\phi)^2}{2\pi Tr^2}\right]\right]\,,
\end{eqnarray}

It is known that $t$ will be the Schwinger parameter up to rescaling by
$T$ if path integral is regarded as the integration of the heat-kernel
method.\cite{7,9} The results agree with the fact that the even modes
correspond to the fields in the adjoint representation of $SO(N)$ and
the odd modes correspond to the fields in the symmetric representation
of 
$SO(N)$; for each even mode, the shifts in the additional masses,
expressed in the unit of $\phi$ here, are proportional to commutator of
the vacuum gauge field and the corresponding field, while for each odd
mode, those are proportional to anticommutator. The massless mode,
which belongs to even modes, corresponds to the excitation of gauge
fields. Then the gauge symmetry is broken to as $SO(N)\rightarrow
SO(N-2)\times U(1)$ in the case considered here.

To summarize:
we find that
the correct mass spectra  are
obtained, through compactification, by means of the path integral
expression with the world-sheet action for open strings which
contains the non-zero background gauge field.

The result should be applied to computation of vacuum  and free
energy of open strings at one-loop level.
In particular, it provides a useful relation in the investigation of
the temperature-dependence of the energy.
In field theory, the free energy
obtained in Ref.~\cite{5} is expressed just in the expansion in terms of
the trace of the Wilson loops. Finite temperature effect only
modifies the ``weight'' for each Wilson loop. But the dominance
of the Wilson loop of the lowest winding number does not undergo a
change in a simple model. Thus we suspect that finite
temperature effect does not have much influence on the Wilson-loop
breaking in general.

As for the case of open strings, Wilson loops are directly
incorporated in the calculation of the path integral. We can treat
the symmetry breaking by Wilson loops in a lucid style.  It is
interesting to consider Wilson loop mechanism in the superstring.
In supersymmetric theories, quantum effects are often cancelled
between bosons and fermions, then the expansion with respect to
Wilson loops might bring about trivial results. The
supersymmetry is broken at finite temperature. The connection between
the evolution of the hot universe and the symmetry breaking may be a
matter of interest.

The study of Wilson loops in the heterotic strings \cite{12} and the
fermionic string theories \cite{13} is required urgently.  We hope to
report on these subjects elsewhere.

\section*{Acknowledgments}
I would like to thank S. Saito for reading this manuscript. I would
also like to thank Iwanami F\=ujukai for financial support.


\end{document}